\documentclass[12pt]{article}
\usepackage{amsfonts}

\begin{document}

\title{Noncommutative Gravity in Six Dimensions}
\author{Cemsinan Deliduman \thanks{%
cemsinan@gursey.gov.tr} \\
\textit{Feza G\"{u}rsey Institute, \c{C}engelk\"{o}y 34684, \.{I}stanbul,
Turkey}}
\date{July, 2006}
\maketitle

\begin{abstract}
A gauge theory of gravity is defined in 6 dimensional non--commutative
space--time. The gauge group is the unitary group $U(2,2)$, which contains
the homogeneous Lorentz group, $SO(4,2)$, in 6 dimensions as a subgroup. It
is shown that, after the Seiberg--Witten map, in the corresponding theory
the lowest order corrections are first order in the non--commutativity
parameter $\theta $. This is in contrast with the results found in
non--commutative gauge theories of gravity with the gauge group $SO(d,1) $.
\end{abstract}

\thispagestyle{empty}

\newpage

\section{Introduction}

Non--commutative (NC) space--times in which the space and time coordinates
do not commute with each other,%
\begin{equation}
\left[ x^{\mu },\,x^{\nu }\right] =i\theta ^{\mu \nu }\left( x\right) \,,
\label{NC}
\end{equation}%
unlike in an ordinary space--time, have caught the imagination of physicists
since Heisenberg. In 1947 Snyder published the first papers that discussed
physics in such space--times \cite{S47}. Since then there has been several
interesting ideas on how to define physics on manifolds with NC geometry.
This interest in defining gauge and gravity theories in NC space--times has
considerably intensified in recent years since the seminal paper of Seiberg
and Witten \cite{SW99}. Although the formulation of non--commutative gauge
theories is well established \cite{JMSSW01}, the Einstein's theory of
gravity or other gravity theories proved to be much more difficult to be
deformed into versions in NC space--times. These so called NC Gravity
theories suffer from the problem of not knowing how to deform general
coordinate invariance and Lorentz symmetry of gravity theories into NC
space--times. There has been several different approaches over the past,
both before the paper of Seiberg and Witten and the after (see \cite{S06}
and references therein). One approach of Chamseddine \cite{C03} is to define
gravity as a gauge theory of a large group (this is $U(2,2)$ in \cite{C03})
which contains the inhomogeneous Lorentz group, $ISO(3,1)$, of general
relativity as a subgroup. This $U(2,2)$ group is then broken into $%
SL(2,C)\times SL(2,C)$ by imposing some constraints and this way one obtains
a version of NC gravity in 4 dimensions. One other approach is to twist the
diffeomorphism symmetry of general relativity \cite{ABDMSW05} and write the
transformation rules in the enveloping algebra. However as noted in \cite%
{AMV06} the so called twisted transformations are not bona fide physical
symmetries in the sense that one cannot derive Ward identities or Noether
currents from the action by using them. However, the star--gauge
transformations, which are obtained by simply deforming commutative theory
gauge transformation rules by insertion of star products, are bona fide
physical symmetries in the sense described above. Therefore a gauge theory
of gravity might be an answer to above mentioned problem. Such a theory in 4
dimensions is described already by Calmet and Kobakhidze in \cite{CK05}.
There in order to escape from the non-invariance issue of $\left[ x^{\mu
},x^{\nu }\right] =i\theta ^{\mu \nu }$ under diffeomorphisms (since $\theta
^{\mu \nu }$ is a constant tensor) the NC version of unimodular theory of
gravity \cite{Uni} is advertised. One common feature of all the above
theories is that the NC\ gravity action can be written as the usual
Einstein--Hilbert action plus infinite number of corrections in a power
series of $\theta $. The NC gravity theory of Calmet and Kobakhidze \cite%
{CK05} is based on gauging inhomogeneous Lorentz group in 4 dimensions and
as noted by Mukherjee and Saha in \cite{MS06} the first order corrections in
$\theta $ in that theory vanish. Actually Mukherjee and Saha's work has more
general conclusions: in any dimensions with a single time coordinate, a NC
gravity theory based on gauging inhomogeneous Lorentz group, $ISO(d,1)$, the
first order corrections will vanish. The second order corrections in $\theta
$ as calculated in \cite{CK06} have very complicated structure. This is a
similar situation in other approaches too: the simple first order correction
terms in $\theta $ vanish and complicated second order corrections survive.

Therefore cosmological implications of these NC gravity theories are hard to
determine. The complexity of the lowest order correction terms in $\theta $
to the general relativity field equations prevents one to seek exact
solutions or to derive modified Friedmann equations. In this paper we are
going to report a new construction of NC gravity which has lowest NC\
corrections to the Einstein's equations proportional to $\theta $ and
moreover these corrections are relatively less complex than the
non-vanishing lowest order\ $\theta ^{2}$ corrections in other NC\ gravity
theories. This theory will thus be better suited to be used to determine the
cosmological implications of a NC\ gravity theory.

Since the NC\ gauge theories are much more well defined than the NC versions
of general relativity, we are going to define the NC gravity theory as a
gauge theory of gravity in the way of Utiyama\cite{U56}. However, as will be
explained below, the correction terms contain the d--coefficients of the
anti--commutation relations between the generators of the gauge group. If
one requires the gauge algebra to be closed also in the anti--commutation
relations, then the gauge group can only be a unitary group. In the
formulation of general relativity by Utiyama one gauges the homogeneous
Lorentz group. The unique case in which the homogeneous \textquotedblleft
special orthogonal\textquotedblright\ Lorentz group of a Lorentzian
space--time is isomorphic to a special unitary group is the Lorentz group in
$4$ space and $2$ time dimensions:%
\begin{equation}
SO(4,2)\cong SU(2,2)\,.
\end{equation}%
However, anti--commutation relations of the generators of $su(2,2)$ do not
close in the algebra. But the generators of $u(2,2)$ do. Therefore the group
that we are going to gauge is $U(2,2)$ in 6 dimensions. Gauging a group
larger than the inhomogeneous Lorentz group might seem going out of
Utiyama's formalism at first, however, as will be seen, the contributions
coming from the $U(1)$ part to the action and the equations of motion will
all vanish. It should be pointed out that our construction is different than
the one by Chamseddine. In \cite{C03} Chamseddine gauges the unitary group
in 4 dimensions and immediately breaks the gauge group into $SL(2,C)\times
SL(2,C) $ by imposing some constraints. Here we describe the theory in 6
dimensions, we do not impose any constraints and also the form of the action
is different. As with the diffeomorphism symmetry we are going to adopt the
prescription given in \cite{CK05}. That is the NC gravity theory, that we
will be describing, will be a unimodular gauge theory of gravity. The NC
gravity theory described in this paper can be reduced to 4 dimensions either
by two time physics \cite{BDA} methods as in \cite{B00} or by considering a
brane--world model in codimension 2. However, since there is an extra time
dimension among the extra coordinates, the codimension 2 brane--world
scenarios would need to be modified accordingly.

\section{NC Gravity Theory in 6 Dimensions}

There are several different ways to write the Einstein's theory of gravity
as a gauge theory \cite{U56}\cite{MM77}\cite{SW80}. One can gauge either the
homogeneous or inhomogeneous Lorentz groups, or a gauge group which contains
either of these. In the latter situation one then imposes some constraints
to obtain the Einstein--Hilbert action. In this paper we will use the
Utiyama's formulation \cite{U56} in which one gauges the homogeneous Lorentz
group. We will also not impose the torsionless condition beforehand, and
therefore we will let the spin connection to be a independent field until
the end. Then we will see that even though the NC torsion vanishes, the
commutative torsion does not. Having the spin connection as an independent
field is called Hilbert--Palatini formalism \cite{P94}. The
Hilbert--Palatini action in 6 dimensions can be written as%
\begin{eqnarray}
S_{U} &=&-\frac{1}{2\kappa _{6}^{2}}\int
d^{6}x\,e\,e_{a}^{m}\,e_{b}^{n}\,R_{mn}\,^{ab}(\omega ) \\
&=&-\frac{1}{96\kappa _{6}^{2}}\int d^{6}x\,\epsilon ^{m_{1}\cdots
m_{6}}\,\epsilon _{a_{1}\cdots
a_{6}}\,e_{m_{1}}^{a_{1}}\,e_{m_{2}}^{a_{2}}\,e_{m_{3}}^{a_{3}}%
\,e_{m_{4}}^{a_{4}}\,R_{m_{5}m_{6}}\,^{a_{5}a_{6}}(\omega )\,,
\end{eqnarray}%
where indices from the middle of the alphabet, $m_{i}=1,\ldots ,6$, are for
the curved space coordinates and indices from the beginning of the alphabet,
$a_{i}=1,\ldots ,6$, are for the tangent space coordinates. Here $\kappa
_{6} $ is related to the $6$ dimensional Newton's constant and the Planck
mass by $\kappa _{6}^{2}=8\pi G_{6}=M_{6}^{-4}$. The Riemann tensor in terms
of spin connection is given by%
\begin{equation}
R_{mn}\,^{ab}(\omega )=\partial _{m}\omega _{n}\,^{ab}-\partial _{n}\omega
_{m}\,^{ab}+\omega _{m}\,^{ac}\omega _{nc}\,^{b}-\omega _{n}\,^{ac}\omega
_{mc}\,^{b}\,.
\end{equation}%
In Utiyama formalism one takes the spin connection as the gauge field of the
homogeneous Lorentz group. We would like to deform this action into a NC\
gravity action in the standard way. That is we are going to replace each
field with its NC counterpart and also replace each product with a star
product. The star product that we are going to use is the Moyal product with
$\theta ^{mn}$ constant. However, as we will comment in the last section,
one can use more general forms of star product.

So we write the NC\ gravity action as%
\begin{equation}
S_{NC}=-\frac{1}{96\kappa _{6}^{2}}\int d^{6}x\,\epsilon ^{m_{1}\cdots
m_{6}}\,\epsilon _{a_{1}\cdots a_{6}}\,\hat{e}_{m_{1}}^{a_{1}}\star \hat{e}%
_{m_{2}}^{a_{2}}\star \hat{e}_{m_{3}}^{a_{3}}\star \hat{e}%
_{m_{4}}^{a_{4}}\star \hat{R}_{m_{5}m_{6}}\,^{a_{5}a_{6}}(\hat{\omega})\,,
\label{action0}
\end{equation}%
where%
\begin{equation}
\star =e^{\frac{i}{2}\overleftarrow{\partial }_{m}\,\theta ^{mn}\,%
\overrightarrow{\partial }_{n}}\,.  \label{Moyal}
\end{equation}

Now we would like to write this theory in terms of a commutative theory to
see what corrections, if any, non-commutativity of space--time brings to the
field equations of general relativity. Since we are gauging only the
homogeneous Lorentz group, we might as well take
\begin{equation}
\hat{e}_{m}^{a}=e_{m}^{a}\,.  \label{vierbein}
\end{equation}

However, the gauge field in the NC theory should be related to the gauge
field in the corresponding commutative theory through Seiberg--Witten map,
which requires the gauge orbits in NC theory and the commutative theory to
be the same. The Seiberg--Witten map written for the spin connection is%
\begin{equation}
\hat{\omega}_{m}\,^{ab}(\omega )+\delta _{\hat{\lambda}}\hat{\omega}%
_{m}\,^{ab}(\omega )=\hat{\omega}_{m}\,^{ab}(\omega +\delta _{\lambda
}\omega )\,.  \label{SWmap}
\end{equation}%
Infinitesimal transformation of commutative field $\omega _{m}\,^{ab}$ is
given by%
\begin{equation}
\delta _{\lambda }\omega _{m}\,^{ab}=\partial _{m}\lambda ^{ab}+\omega
_{m}\,^{ac}\lambda _{c}\,^{b}-\omega _{m}\,^{bc}\lambda _{c}\,^{a}\,,
\end{equation}%
For the deformed field one assumes the same form, however with every
commutative theory object is exchanged with the corresponding NC theory
object as follows
\begin{equation}
\delta _{\hat{\lambda}}\hat{\omega}_{m}\,^{ab}=\partial _{m}\hat{\lambda}%
^{ab}+\hat{\omega}_{m}\,^{ac}\star \hat{\lambda}_{c}\,^{b}-\hat{\lambda}%
^{ac}\star \hat{\omega}_{mc}\,^{b}\,.
\end{equation}%
In order to find a solution to eq.(\ref{SWmap}), in \cite{JMSSW01, C05} both
the gauge field and the gauge transformation parameter are written in a
series expansion in the deformation parameter $\theta $ as%
\begin{equation}
\hat{\omega}_{m}\,^{ab}=\omega _{m}\,^{ab}+\omega
_{m}^{^{(1)}}\,^{ab}(\omega )+O\left( \theta ^{2}\right) \,,
\end{equation}%
\begin{equation}
\hat{\lambda}^{ab}=\lambda ^{ab}+\lambda ^{^{(1)}\,ab}(\lambda ,\omega
)+O\left( \theta ^{2}\right) \,,
\end{equation}%
where $\omega _{m}^{^{(1)}}\,^{ab}(\omega )$ and $\lambda
^{^{(1)}\,ab}(\lambda ,\omega )$ are order $\theta $ quantities. Then a
solution to eq.(\ref{SWmap}) to first order in $\theta $ is found to be \cite%
{JMSSW01, C05}
\begin{equation}
\hat{\omega}_{m}\,^{ab}=\omega _{m}\,^{ab}-\frac{i}{4}\theta ^{kl}\left\{
\omega _{k},\partial _{l}\omega _{m}+R_{lm}\right\} ^{ab}+O\left( \theta
^{2}\right) \,.  \label{NCsc}
\end{equation}%
The deformed Riemann tensor is given by the usual prescription,%
\begin{equation}
\hat{R}_{mn}\,^{ab}(\hat{\omega})=\partial _{m}\hat{\omega}%
_{n}\,^{ab}-\partial _{n}\hat{\omega}_{m}\,^{ab}+\hat{\omega}%
_{m}\,^{ac}\star \hat{\omega}_{nc}\,^{b}-\hat{\omega}_{n}\,^{ac}\star \hat{%
\omega}_{mc}\,^{b}\,.
\end{equation}%
Up to first order in $\theta $ this can be easily calculated:%
\begin{eqnarray}
\hat{R}_{mn}\,^{ab} &=&R_{mn}\,^{ab}+\mathring{R}_{mn}\,^{ab} \\
\mathring{R}_{mn}\,^{ab} &=&\frac{i}{2}\theta ^{kl}\left[ \left\{
R_{mk},R_{nl}\right\} ^{ab}-\frac{1}{2}\left\{ \omega _{k},\left( \partial
_{l}+\nabla _{l}\right) R_{mn}\right\} \right] \\
&=&\frac{i}{2}\theta ^{kl}R_{mk}\,^{cd}R_{nl}\,^{ef}d_{cd,ef}\,^{ab}-\frac{i%
}{4}\theta ^{pr}\omega _{k}\,^{cd}\left( \partial _{l}+\nabla _{l}\right)
R_{mn}\,^{ef}d_{cd,ef}\,^{ab}\,,  \label{defRiem1}
\end{eqnarray}%
where $d_{cd,ef}\,^{ab}$ are the d--coefficients of the gauge group. We name
$\mathring{R}_{mn}\,^{ab}$ as the deformed curvature tensor. Plugging these
expression into the NC gravity action, one obtains up to the first order in $%
\theta $
\begin{eqnarray}
S_{NC} &=&-\frac{1}{96\kappa _{6}^{2}}\int d^{6}x\,\epsilon ^{m_{1}\cdots
m_{6}}\,\epsilon _{a_{1}\cdots
a_{6}}\,e_{m_{1}}^{a_{1}}\,e_{m_{2}}^{a_{2}}\,e_{m_{3}}^{a_{3}}%
\,e_{m_{4}}^{a_{4}}\,\left( R_{m_{5}m_{6}}\,^{a_{5}a_{6}}(\omega )+\mathring{%
R}_{m_{5}m_{6}}\,^{a_{5}a_{6}}(\omega )\right) \\
&=&-\frac{1}{2\kappa _{6}^{2}}\int d^{6}x\,e\,e_{a}^{m}\,e_{b}^{n}\,\left(
R_{mn}\,^{ab}(\omega )+\mathring{R}_{mn}\,^{ab}(\omega )\right) \\
&=&-\frac{1}{2\kappa _{6}^{2}}\int d^{6}x\,e\,\left( R(\omega )+\mathring{R}%
(\omega )\right) \,,  \label{action}
\end{eqnarray}%
where we have made the definition%
\begin{equation}
\mathring{R}(\omega )=e_{a}^{m}\,e_{b}^{n}\,\mathring{R}_{mn}\,^{ab}(\omega
)\,.  \label{defRicciSca}
\end{equation}%
We are going to call this quantity as the deformed curvature scalar.

Next we would like to calculate the d--coefficients of the gauge group $%
U(2,2)$ and show that not all parts of the above NC correction term, i.e.
deformed curvature scalar, vanish. In order to calculate the d-coefficients
of anti-commutation relations of the algebra $u(2,2)$ we used the isomorphy $%
u(2,2)\,\cong \,Cl\left( 2,2\right) \,\cong \,o(4,2)$, where the $Cl\left(
2,2\right) $ is the Clifford algebra, generated by 4 gamma matrices \cite%
{C05}\cite{vN81}. Denoting the $so(4,2)$\ generators as $\gamma _{ab}$ one
can write their commutation and anti--commutation relations from the
corresponding relations in $u(2,2)\,\cong \,Cl\left( 2,2\right) \,$. $\gamma
_{ab}$ obey the Lorentz algebra in 6 dimensions:%
\begin{equation}
\left[ \gamma _{ab},\gamma _{cd}\right] =2\eta _{ad}\,\gamma _{bc}+2\eta
_{bc}\,\gamma _{ad}-2\eta _{ac}\,\gamma _{bd}-2\eta _{bd}\,\gamma _{ac}\,,
\end{equation}%
where $\eta _{ab}=sign\left( +,-,-,-,+,-\right) $. One can write the
anti--commutation relations and calculate the d--coefficients from the
corresponding expressions in $u(2,2)\,\cong \,Cl\left( 2,2\right) $. We find
them as%
\begin{equation}
\left\{ \gamma _{ab},\gamma _{cd}\right\} =2i\epsilon _{abcd}\,^{ef}\,\gamma
_{ef}+2\left( g_{ad}\,g_{bc}-g_{ac}\,g_{bd}\right) \mathbb{I}\,.
\end{equation}%
It is in this relations that one sees the need to gauge $u(2,2)\,\cong
\,o(4,2)$, but not $su(2,2)\,\cong \,so(4,2)$. The $u(1)$ part explicitly
appears on the right hand side. Plugging the expression for d-coefficients
into (\ref{defRiem1}) we find the deformed Ricci scalar (\ref{defRicciSca})
as
\begin{equation}
\mathring{R}\,(\omega )=-\theta ^{kl}e\,\epsilon ^{prstuv}\left(
R_{pkst}\,R_{rluv}-\frac{1}{2}\omega _{kst}\,\left( \partial
_{l}+D_{l}\right) R_{pruv}\right) -2\theta ^{kl}\left(
R_{mk}\,^{ab}\,R^{m}{}_{l\,ab}\right) \,.
\end{equation}%
By using the anti-symmetry of $\theta ^{kl}$ tensor and the symmetry
properties of the Riemann tensor as in \cite{MS06}, it can be shown that the
last term vanishes. Thus, up to order $\theta $, the NC correction term is%
\begin{equation}
\mathring{R}\,(\omega )=-\theta ^{kl}e\,\epsilon ^{prstuv}\left(
R_{pkst}\,R_{rluv}-\frac{1}{2}\omega _{kst}\,\left( \partial
_{l}+D_{l}\right) R_{pruv}\right) \,.  \label{defRicciSca2}
\end{equation}

To obtain the deformed Einstein field equations we plug (\ref{defRicciSca2})
into the action (\ref{action}) and then vary it with respect to $e_{a}^{m}$.
We obtain%
\begin{equation}
R_{mn}(\omega )+\mathring{R}_{mn}(\omega )-\frac{1}{2}\left( R(\omega )+%
\mathring{R}(\omega )\right) g_{mn}=0  \label{eom}
\end{equation}%
as the deformed Einstein field equations in vacuum. Here we also defined a
deformed Ricci tensor as%
\begin{eqnarray}
\mathring{R}_{mn}\,(\omega ) &=&\mathring{R}_{mr}\,^{ab}(\omega
)\,e_{b}^{r}\,e_{na} \\
&=&-\theta ^{kl}e\,\epsilon _{m}\,^{rstuv}\left( R_{nkst}\,R_{rluv}-\frac{1}{%
2}\omega _{kst}\,\left( \partial _{l}+D_{l}\right) R_{nruv}\right) \,.
\end{eqnarray}%
The field equation (\ref{eom}) has a very simple form and since
the NC correction terms $\mathring{R}_{mn}(\omega )$ and
$\mathring{R}(\omega )g_{mn}$ are first order in $\theta $, they
are also not very complicated. We note that the existence of these
non--trivial first order $\theta $ corrections to equations of
motion in our NC gravity theory is due to the fact that we are
gauging the $SO(4,2)$ group, but not the $SO(5,1)$ group.

The equation (\ref{eom}) is the main result of this paper. But before
commenting on this equation we would like to discuss the torsion induced due
to the non--commutativity of the coordinates. As it is stated we work in the
Hilbert--Palatini formalism and therefore we treat the spin connection
independent of the vierbein degrees of freedom. Then we need also to vary
the action (\ref{action0}) with respect to the NC spin connection. This
variation gives us the \textquotedblleft non--commutative\textquotedblright\
torsionless condition%
\begin{equation}
\hat{D}_{[m}\,\hat{e}_{n]}^{a}=0\,,
\end{equation}%
which is, due to (\ref{vierbein}), equivalent to%
\begin{equation}
\hat{D}_{[m}\,e_{n]}^{a}=\partial _{\lbrack m}\,e_{n]}^{a}+\hat{\omega}%
_{[m}\,^{ab}\,e_{n]b}=0\,.
\end{equation}%
Writing the spin connection of NC theory in terms of spin connection of
commutative theory up to order $\theta $ (\ref{NCsc}) we find%
\begin{equation}
\partial _{\lbrack m}\,e_{n]}^{a}+\omega _{\lbrack m}\,^{ab}\,e_{n]b}-\frac{i%
}{4}\theta ^{kl}\left\{ \omega _{k},\partial _{l}\omega _{\lbrack
m}+R_{l[m}\right\} ^{ab}\,e_{n]b}=0\,.
\end{equation}%
Then the torsion of the corresponding commutative theory is%
\begin{equation}
T_{mn}\,^{a}=-2D_{[m}\,e_{n]}^{a}=-\frac{i}{2}\theta ^{kl}\left\{ \omega
_{k},\partial _{l}\omega _{\lbrack m}+R_{l[m}\right\} ^{ab}\,e_{n]b}\,,
\end{equation}%
where we used the convention of \cite{vN81}. Therefore there is a
non--vanishing torsion in the commutative theory and it is proportional to
the non--commutativity tensor. In a sense the non--commuting nature of the
coordinates creates this torsion.

\section{Comments and Conclusions}

We summarize the main aspects of the construction presented in this paper
before commenting on the main results. We defined a gauge theory of gravity
in $6$ dimensional space--time with non--commuting coordinates. The reason
that we formulated the gravity theory as a gauge theory is because the
deformations of gauge theories is much better defined than the deformation
of gravity theories. The non-invariance of the constant tensor $\theta ^{\mu
\nu }$ forces one to consider the unimodular theory of gravity in NC
space--time as in \cite{CK05}. We followed the Utiyama's approach in
formulation of gauge theory of gravity and chose the gauge group as the
homogeneous Lorentz group in $6$ dimensions. This group, being $SO(4,2)$, is
isomorphic to the special unitary group $SU(2,2)$. In NC gauge theories,
after the Seiberg--Witten map, the corrections to the gauge field (here the
spin connection) or to the field strength (here the curvature tensor)
contain the d--coefficients of the anti--commutation relations of Lie
algebra generators and those anti--commutation relations close only in the
case of unitary groups. Therefore we gauged $U(2,2)$ instead of $SO(4,2)$.
In fact the reason that we worked in $6$ dimensions after all is due to the
fact that only in $6$ dimensional Lorentzian space--time with $4$ space and $%
2$ time dimensions the homogeneous Lorentz group is isomorphic to a unitary
group. For space--times whose homogeneous Lorentz group is $SO(d,1)$ it is
already shown \cite{MS06} that the first order corrections in $\theta $ of
NC\ gravity to Einstein--Hilbert action vanish. In contrast in our
construction those corrections do not vanish. Since in the other
constructions the second order corrections in $\theta $ of NC\ gravity to
the Einstein--Hilbert action or to the field equations are very complicated,
there is little hope to see the effects of those correction terms in a
cosmological setting. However, the first order corrections in our case are
relatively simple and might have some relevance on the cosmological problems.

To make contact with cosmology and to assess the effects of
non--commutativity of coordinates on cosmology through NC\ gravity one needs
to reduce the theory, that we described in $6$ dimensions, into $4$
dimensions. This can be done in several ways. Since this theory is described
in a space--time with two time dimensions, one way is to use the techniques
of two--time physics \cite{BDA} as described in \cite{B00} (see also \cite%
{VP99} for similar ideas). Having two time--like directions in the
space--time might seem, at first look, unphysical. To make sense
of such a theory one needs to derive a \textquotedblleft
one--time\textquotedblright\ theory from the theory with two
time--like dimensions. The basic idea behind the two--time physics
(see \cite{Brev} for reviews of two--time physics) is that the
\textquotedblleft evolution parameter\textquotedblright\ which
will be interpreted as the physical time in lower dimensional
one--time theory is either a gauge choice in the higher
dimensional two--time theory (in the case of particle and
tensionless brane theories), or it is obtained by imposing some
kinematical constraints (in the case of field theory). That is we
do not interpret any of the time--like coordinates in space--time
with two times as evolution parameters. The true evolution
parameter, therefore the physical time, is the coordinate one
obtains as a gauge choice in the subspace of two time and one
extra space dimensions. Time is also a gauge choice in general
relativity and in that sense the treatment of time in two--time
physics is similar. In the case of particle theories, the gauge
symmetry that one uses to reduce a two--time theory to a one--time
theory is the $Sp(2,R)$ symmetry of the phase space, promoted to a
local symmetry. In the case of field theories defined on
space--times with two times, the kinematical constraints, that one
needs to impose in order to obtain a physical field theory with
one time, obey the same $Sp(2,R)$ algebra. The two--time physics
is the only rigorous way to make sense of a theory defined on a
space--time with two time--like coordinates. The problems with
causality and unitarity do not exist in two--time physics, because
the final theory has only one time as the evolution parameter with
a well defined Hamiltonian. Two--time physics reduction of the
present theory will be done in a future publication. The result of
that research is expected to be a non-trivial modification of the
Einstein equations in $4$ dimensional space--time with NC
corrections in the first order in $\theta $.

One other way of reducing the theory to $4$ dimensions is to
consider a brane--world model in codimension $2$ and to find the
induced gravity field equations or the modified Friedmann
equations on the $3-$brane. Here again due to having two
time--like dimensions one cannot construct a conventional
codimension $2$ brane--world model, but should consider embedding
a $3-$brane into a space--time with two times. Isometric embedding
of BPS branes in space--times with two times is analyzed some time
ago by L. Andrianopoli et al. in \cite{ADGHSP}. As it is commented
there the minimal embedding of world--volume geometry of a genuine
BPS brane of string theory in a higher dimensional space--time
requires at least two extra dimensions \cite{Eisenhart} and that
the higher dimensional space--time has to have at least two
time--like dimensions \cite{Penrose}. Therefore $6$ dimensional
space--time with two times is the minimal choice to embed a BPS
$3-$brane of string theory \cite{ADGHSP}. Done either with the
methods of two--time physics or by embedding a BPS $3-$brane in
the $6$ dimensional space--time with two times, the main aim of
reducing the theory to $4$ dimensions would be first to determine
what modifications of the NC gravity theory described in this
paper will survive in $4$ dimensions and then how this
modifications of Einstein equations will modify the Friedmann
equations and therefore the evolution of the universe.
Specifically we would like to understand whether the
non--commutativity of the coordinates and the NC gravity has
anything to say about the still unsolved dark matter and dark
energy problems. Modified Friedmann equations could be the first
step in the direction of such an understanding \cite{Mann}.

Defining NC gravity theory in a higher dimensional space--time opens up also
exciting new possibilities. Now it is possible to have just the extra
dimensions non--commutative and therefore get rid of all the problems
created in $4$ dimensions by the non--commuting nature of the coordinates.
In such a scenario, since $4$ dimensional space--time would be an ordinary
space--time with commuting coordinates, the breaking of Lorentz invariance
in $4$ dimensions could be avoided and stringent bounds \cite{CHK01}\cite%
{JLM01} on the value of the non--commutativity parameter $\theta $ could be
lifted.

As it is commented before eq.(\ref{action0}) the star product need not to be
the Moyal product (\ref{Moyal}). More general forms of star products, even
with coordinate dependent $\theta ^{mn}$, can be used. For example one can
try a construction similar to the one described in \cite{GGR05} by using the
Rieffel product \cite{R93}. In this case, it is again possible to restrict
the non--commutativity into just the extra dimensions. Then the only
non-zero components of $\theta ^{mn}$ will be $\theta ^{56}=-\theta ^{65}$.
These components may be made to depend on 4D coordinates on the $3-$brane
and consequences of this position dependent non--commutativity can be
analyzed. Works in the mentioned lines of research are still in progress.
The NC gravity theories in higher dimensions have many promising avenues of
research and it will be exciting to see whether they will help us to answer
some of the profound questions in $4$ dimensional cosmology.

\end{document}